# Electron energy loss spectroscopy with parallel readout of energy and momentum

Harald Ibach
*Peter Grünberg Institut (PGI-6), Forschungszentrum Jülich, 52425 Jülich, Germany*

François C. Bocquet[*], Jessica Sforzini, Serguei Soubatch, F. Stefan Tautz
*Peter Grünberg Institut (PGI-3), Forschungszentrum Jülich, 52425 Jülich, Germany*
*Jülich Aachen Research Alliance (JARA), Fundamentals of Future Information Technology, 52425 Jülich, Germany*

We introduce a high energy resolution electron source that matches the requirements for parallel readout of energy and momentum of modern hemispherical electron energy analyzers. The system is designed as an add-on device to typical photoemission chambers. Due to the multiplex gain, a complete phonon dispersion of a Cu(111) surface was measured in seven minutes with 4 meV energy resolution.

## I. INTRODUCTION

It is now almost 50 years ago that Propst and Piper published the first spectra of electron energy losses caused by vibrational excitations on a W(100) surface [1]. Since then, the performance of spectrometers has improved greatly with resolutions reaching down to 0.5 meV [2,3]. Instrumental for the improvement was the invention of a new type of electrostatic deflector which features active stigmatic focusing at 146° deflection angle and angular aberration correction [2]. Because of the active stigmatic focusing this monochromator can carry large current loads without space charge induced aberrations and is therefore the best choice for producing intense highly monochromatic beam of electrons.

Electron energy loss spectroscopy (EELS) and in particular the high resolution EELS (HREELS) was successfully employed in studies of localized vibrations of adsorbed species, surface phonons and plasmons (for an overview see Ref. [4]), and recently also magnons [5-8]. Probing magnons proved to be particularly demanding since the scattering probability d$P$/d$\Omega$ is only of the order of $10^{-5}$, i.e., nearly two orders of magnitude lower than the probability for phonon scattering [9,10] and several orders of magnitude lower than the probability of inelastic scattering from dipole active modes such as the stretching vibration of adsorbed carbon monoxide [11]. In the latter case the electron interacts with long-range dipole fields associated with vibrational excitations. Because of the long range nature of the interaction the inelastic scattering is focused in the direction of the specular reflected beam (and diffracted beams); in other words only the center of the surface Brillouin zone is probed. Because of the high intensity the full information on dipole active energy losses is obtained in a few minutes when using a conventional single channel spectrometer.

In the case of phonon and magnon scattering, the dispersion of the excitation energy as function of wave vector transfer is of interest, and the intensities are low. Collecting a complete set of data points sufficient to describe the phonon or magnon dispersion in the conventional sequential mode typically requires a day's work for the most advanced single channel spectrometers [12,13], or several days with a conventional spectrometer. The situation is aggravated by the fact that cross sections for phonon and magnon scattering depend strongly on the electron impact energy with the consequence that experiments must include the search for an optimum value of the impact energy. A parallel detection of electrons of different loss energy and angle would therefore be highly desirable. Parallel detection of electrons of different kinetic energies travelling in different directions within some (*acceptance*) angle is nowadays the standard operation mode of hemispherical deflector analyzers, for example in the analysis of photoemitted electrons. These analyzers use the two-dimensional optical readout of a multichannel plate (MCP).

*f.bocquet@fz-juelich.de

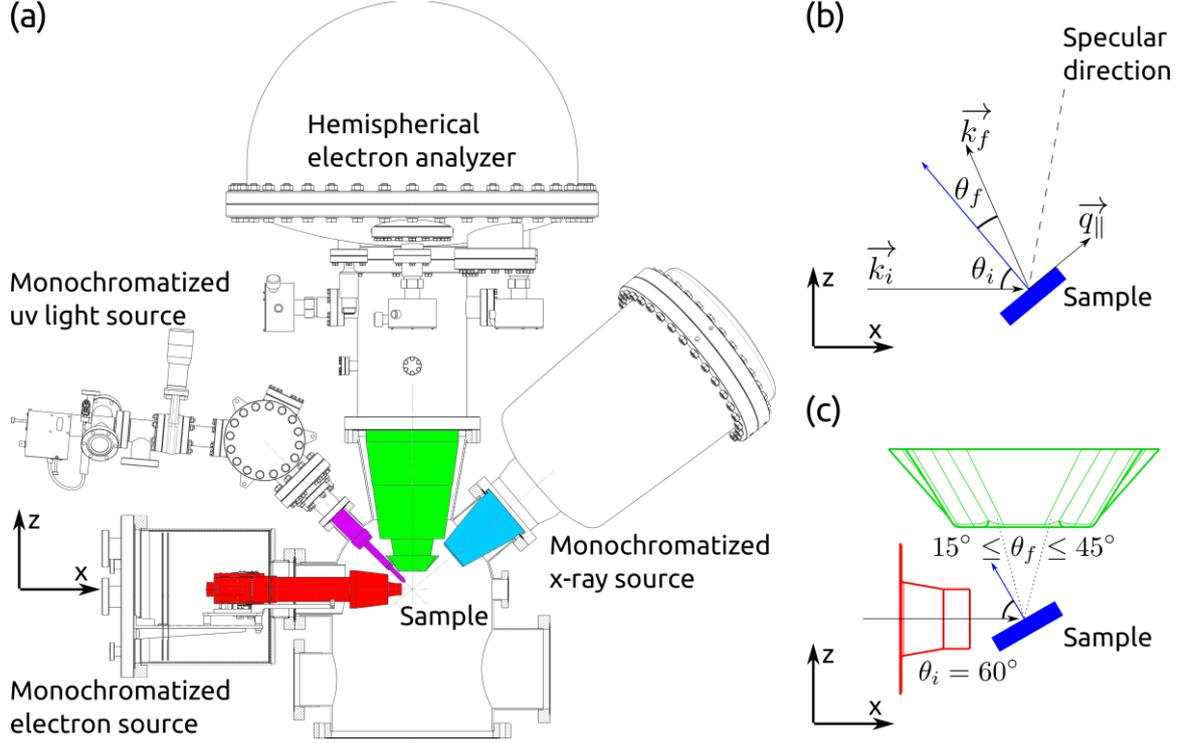

Fig. 1. (a) Sketch of the photoemission chamber equipped with the monochromatic electron source (red). The chamber and the electron source chamber are cut in the *xz*-plane. The optical axes of the hemispherical analyzer lens system, of the photon sources (ultraviolet light and x-ray), of the electron source and the long axes of the electron source exit slit and of the analyzer entrance slit are all in the same *xz*-plane, the scattering plane. (b) Scattering geometry and definition of vectors and angles. (c) Geometry for $\theta_i = 60°$. The dashed lines represent the ±15° angular acceptance of the analyzer.

It is therefore a natural thought to combine the use of 146° deflectors in the monochromatic electron source with a hemispherical 180° electron analyzer featuring parallel detection. Such a combination was recently reported by Xuetao Zhu *et al.* [14] with a dedicated apparatus design.

In this publication, we describe a high energy resolution electron source that is designed to be used as an add-on-instrument to commercially available photoemission vacuum chambers equipped with a hemispherical analyzer.

## II. BASIC CONSIDERATIONS

Modern hemispherical analyzers with a two-dimensional readout designed for photoemission experiments typically feature two main modes of operation: the *transmission* mode and the *angular* mode. The main difference between these modes is that in the *transmission* mode, the lens system of the analyzer focuses electrons emitted from different positions along a line on the sample (independent of the emission direction) on different positions along the analyzer entrance slit, while in the *angular* mode it does the same for electrons emitted in different directions (independent of the emission position). Accordingly, electrons passing through the entrance slit and further dispersing between the hemispheres arrive at the MCP detector such that in the radial direction one obtains the energy dispersion (typical range of about 8% of the analyzer pass energy), while in the azimuthal direction either spatial or angular dispersion is delivered for the *transmission* or the *angular* mode, respectively. For a sample possessing a crystalline symmetry the angular resolution capability of the analyzer is equivalent to the wave vector (or to the *reciprocal* space) resolution.

Fig. 1a shows our experimental set-up. The ultra-high vacuum (UHV) chamber is equipped with monochromatized sources of ultraviolet and x-ray radiation and a hemispherical electron analyzer (Scienta R4000) in the typical configuration of a modern photoemission apparatus, but with an additionally installed monochromatic electron source, controlled by a Scienta power supply and a home-made software. In order to make the monochromatic electron source as easily mountable as a light source, the cathode emission system and the double monochromator, consisting of two 146° deflectors, are located in a small independent chamber that is bolted to the main photoemission chamber via a CF-150 flange, as depicted in Fig. 1a. To bridge the distance of 220 mm between the connecting flange and the center of the analysis chamber, a set of lenses (*transfer lens*) is designed. The distance between the exit slit of the second monochromator and the sample is 300 mm. Both the electron source chamber and the analysis chamber possess a double-wall μ-metal magnetic shielding. The residual magnetic field in both chambers and at the connecting flange is below 2 mG.

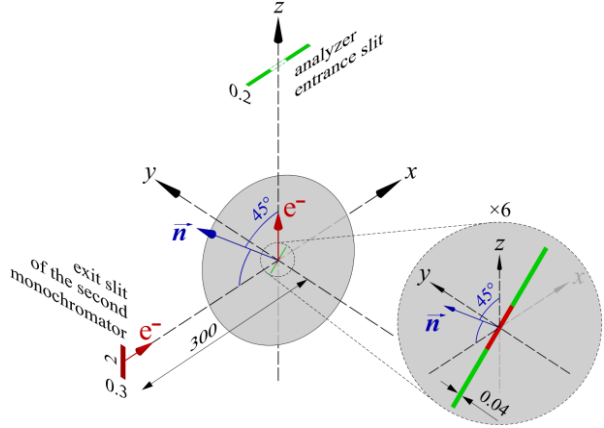

Fig. 2. Schematic view of the electron scattering geometry. The exit slit of the second monochromator and the electron beam spot at the sample are shown in red. The entrance slit of the hemispherical analyzer and the maximal illuminated area that can be viewed by the analyzer to keep optimum performances are shown in green. The sample has a tilt of 45° from the optical axis of the analyzer lens system toward the optical axis of the electron source transfer lens.

The design of the two deflectors is as described in Ref. [12, 13]. The first deflector (mono1) runs with about 3-10 times the deflection voltage of the second deflector (mono2) in order to reduce space charge effects. The deflection voltage of the second deflector ($\Delta V_{\text{mono2}}$) determines the electron beam energy resolution. The relation between $\Delta V_{\text{mono2}}$ and the theoretical energy resolution $\Delta E_{\text{mono2}}^{\text{theo}}$ (full width at half maximum, FWHM) is

$$\Delta E_{\text{mono2}}^{\text{theo}} \approx 2.5 \times 10^{-3} e \times \Delta V_{\text{mono2}}, \quad (1)$$

where $e$ is the elementary charge. The exact energy resolution also depends on the spread of angles of the electrons feeding the monochromator. The relation between the pass energy $E_{\text{mono2}}^{\text{pass}}$ (the energy of electrons travelling along the center path) and the deflection voltage $\Delta V_{\text{mono2}}$ is

$$E_{\text{mono2}}^{\text{pass}} = 0.54 e \times \Delta V_{\text{mono2}}. \quad (2)$$

The resolution of the hemispherical analyzer depends on the pass energy and the choice of the entrance slit. In test measurements we used the curved (25×0.2) mm² slit. This slit is positioned in the *xy*-plane of Fig. 1, its long axis points in the *x* direction. Since the magnification of the analyzer lens system in the transmission mode equals 5, the *xy*-area at the sample position that is viewed by the analyzer is about (5×0.04) mm². In the angular mode of the analyzer lens system, the area illuminated by electrons should be smaller than (1×0.1) mm² in order to keep angular aberrations small and high intensity [15]. Hence, in order to be compatible with both operation modes of the analyzer, the monochromatic electron source is required to deliver a spot onto the sample surface whose projection into the *xy*-plane is smaller than (1×0.04) mm², see Fig. 2.

Our second monochromator features an exit slit of (0.3×2) mm², oriented in the *yz*-plane in Fig. 1. Having the long axes of the slits of the analyzer and the monochromator in *x*- and *z*-directions, respectively, allows matching the area illuminated by the electron source with the area viewed by the analyzer. To actually achieve this matching, the image of the monochromator slits at the sample must be reduced by a factor of 7.5 to reach the size of (0.04×0.27) mm². Assuming a tilt angle of 45° of the surface normal of the sample against the optical axis of the electron source, the projection into the *xy*-plane of the illuminated spot in the sample surface is (0.27×0.04) mm², see Fig. 2. The extension in the *x*-direction is thus much smaller than the required 1 mm.

The reduced image size entails relatively large angles of the incoming electrons with respect to the optical axis. The spread of angles which the electrons form with the optical axis ultimately stems from the spread of angles in the feed beam of the first monochromator, which in turn is produced by the cathode emission system. Phase-space conservation requires that the spreading of angles of the feed beam produced by the cathode system transfers into the spreading of angles at the target (i.e. the sample). Concerning the angle $\alpha$ in the dispersion plane ($xy$-plane in Fig. 1) of the monochromators the relation reads:

$$s_\alpha^{\text{cath}} w_\alpha^{\text{cath}} \sqrt{E_{\text{mono1}}^{\text{pass}}} = s_\alpha^{\text{target}} w_\alpha^{\text{target}} \sqrt{E_0}. \quad (3)$$

Here, $s_\alpha^{\text{cath}}$ and $s_\alpha^{\text{target}}$ are the variances of the Gaussian distributions of angles delivered by the cathode feed system and at the target, respectively; $w_\alpha^{\text{cath}}$ and $w_\alpha^{\text{target}}$ are the slit widths of the first monochromator and its image at the target; $E_{\text{mono1}}^{\text{pass}}$ and $E_0$ are the kinetic energies of the electrons in the first monochromator and at the target, respectively. A standard mode of operation with high resolution uses $E_{\text{mono1}}^{\text{pass}} \approx 4\,\text{eV}$. The Gaussian variance of angles delivered by the cathode $s_\alpha^{\text{cath}}$ is then $0.6°$. With $w_\alpha^{\text{cath}} = 0.3\,\text{mm}$, $w_\alpha^{\text{target}} = 0.04\,\text{mm}$ and $E_0 = 60\,\text{eV}$ one calculates $s_\alpha^{\text{target}} = 1.16°$. Similar consideration in the plane perpendicular do the dispersion plane (the $xz$-plane) yields $s_\beta^{\text{target}} = 1.72°$, using $w_\beta^{\text{target}} = 3\,\text{mm}$, $w_\beta^{\text{target}} = 0.27\,\text{mm}$ and $s_\beta^{\text{cath}} = 0.6°$. The FWHM of the $\beta$-angle ($4.05°$) thus calculated amounts for a considerable fraction of the $30°$ acceptance angle of the analyzer. However, we will show in Sec. V that the actual transmission of the whole electron source is small for large $\beta$-angles.

We remark in passing that the standard mode of operation of a conventional HREELS spectrometer *does not* involve the formation of an image of the monochromator exit slit at the sample (see Ref. [12] for details). Here, however, image formation is required, since the analyzer requires the illuminated area to be small for optimum performance. Preferably all electrons produced by the monochromatized source should impinge on the surface within that area. That is possible only when the illuminated area is an image of the monochromator exit slit. The momentum range which can be probed by the hemispherical analyzer is determined by the scattering geometry on the one hand, and the range of accepted emission angles on the other. The optical axes of the electron monochromator and the analyzer lens system form an angle of $90°$ (Fig. 1a). The maximum accepted angle relative to the optical axis of the analyzer lens is $\pm 15°$ for the Scienta R4000 analyzer employed here.

The wave vector range viewed by the analyzer is calculated from wave vector conservation:

$$q_\parallel = k_i \sin(\theta_i) - k_f \sin(\theta_f) \quad \text{if} \quad E_f = E_i - \hbar\omega \quad (4)$$

$$-q_\parallel = k_i \sin(\theta_i) - k_f \sin(\theta_f) \quad \text{if} \quad E_f = E_i + \hbar\omega. \quad (5)$$

Here $\hbar\omega$ and $q_\parallel$ are energy and parallel component of the wave vector of the excitation; $\theta_i$ and $\theta_f$ are the angles of incidence and emission with respect to the surface normal; $k_i$ and $k_f$ are the wave vectors of incident and backscattered electrons, respectively (see Fig. 1b). The tilt angle of the sample is typically chosen such that the intense specular reflected beam falls just outside the angle range viewed by the analyzer. For instance, this is the case with an angle $\theta_i$ of $60°$, as depicted in Fig. 1c. In this configuration, the angle $\theta_f$ accepted by the analyzer ranges between $15°$ and $45°$ with respect to the surface normal. For small energy losses the probed wave vector $q_\parallel$ ranges from:

$$|q_\parallel^{\max}| = \sqrt{\frac{2m_e E_0}{\hbar^2}} (\sin 60° - \sin 15°) \quad (6)$$
$$= 3.11\,\text{nm}^{-1} \sqrt{E_0/\text{eV}}$$

to

$$|q_\parallel^{\min}| = \sqrt{\frac{2m_e E_0}{\hbar^2}} (\sin 60° - \sin 45°) \quad (7)$$
$$= 0.82\,\text{nm}^{-1} \sqrt{E_0/\text{eV}}$$

where $m_e$ is the electron mass.

Wave vectors closer to the center of the Brillouin zone can be reached by moving the reflected beam further toward the optical axis of the analyzer lens system. In order to study the wave vector range from the center to the boundary of the Brillouin zone in a single frame, relatively high impact energies of 50-100 eV are required. Such high impact energies are typically used for inelastic scattering from phonons [4, 16]. For inelastic scattering from spin waves, on the other hand, low impact energies between 2-8 eV are required for an optimum cross section [8, 9]. In that case only a small portion of the Brillouin zone is measured.

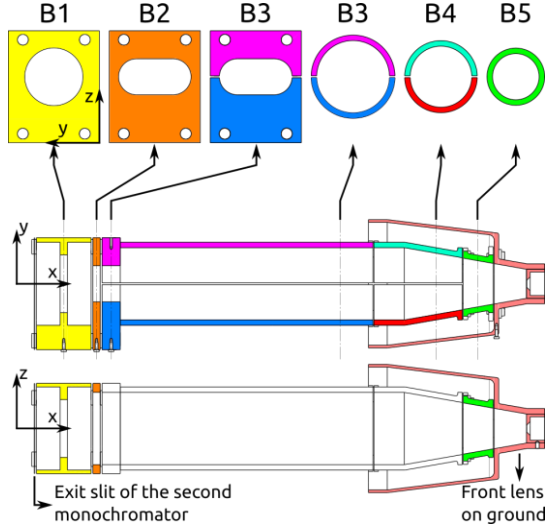

Fig. 3. Design of the transfer lens: side view (bottom), top view (middle) and cross sections of the various electron-optical elements (top).

In the conventional HREELS spectrometer design, the sample potential varies with the impact energy. The sample is surrounded by metal plates, the so-called scattering chamber, kept at the same potential as the sample to provide a zero-field environment in the vicinity of the sample. Commercial hemispherical analyzers do not require scattering chambers, as the sample is typically grounded. The potential of the cathode tip of our electron source is therefore set to negative values with respect to ground by an amount that determines the impact energy $E_0$ of electron at the target. To vary the electron impact energy, all potentials of the cathode and of both monochromators must be varied by the same amount. In practice, this is achieved by referencing all these potentials to a common potential, which is negative with respect to ground. Changing the impact energy then requires only the change of this common potential. Thus, all potential differences from the cathode to the monochromator exit slit remain strictly constant when changing the impact energy. Only the potentials of the transfer lens system need be adjusted, and for this a "*lens table*" can be used, as will be shown in the following section.

## III. THE TRANSFER LENS BETWEEN MONOCHROMATOR AND TARGET

This section describes the design of the transfer lens and its properties. As remarked above, the task of the lens is to form a reduced-size image of the exit slit of the second monochromator at the target for a wide range of impact energies. Reduction of the size by a factor as large as 7.5 in a single step would involve electron trajectories relatively far off the optical axis, which causes large aberrations. Therefore the reduction of the image size was conducted in two steps via an intermediate image.

Fig. 3 shows cross sections of the transfer lens. The lens involves six elements. Two of them, B3 and B4, are split. The reason for the splitting is the following: The only beam parameter of the second monochromator over which one does not have complete control is the mean exit angle in the dispersion plane (*xy*-plane in Fig. 3), since the exit angle changes somewhat with the current load in the first monochromator as well as with the retardation ratio between first and second monochromator. In the conventional HREELS this fact is of little concern because of the short path length between monochromator and target. Here, it matters, however.

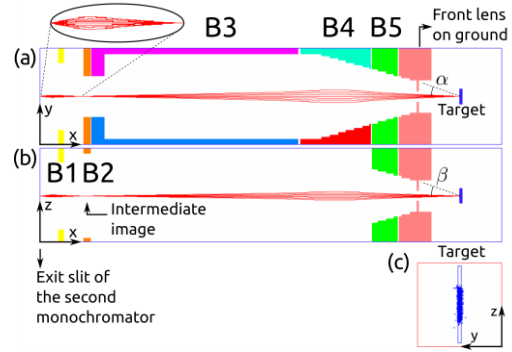

Fig. 4. Focusing of the electron beam in the transfer lens for the case $\alpha_{mean} = 0°$. (a) Top view into the dispersion plane of the monochromators ($\alpha$-plane / *xy*-plane). (b) Side view ($\beta$-plane / *xz*-plane). (c) Projection of the electron beam spot into the *yz*-plane at the target (blue dots) and projection of the field of view of the hemispherical analyzer into the *yz*-plane at the target (blue-lined frame). Electron trajectories, shown in red, are calculated for the following conditions: $V_{B1}=-4.1$ V, $V_{B2}=12.14$ V, $V_0=59.46$ V, $E_0=60$ eV, $V_{B5}=59.46$ V, $V_{B3left}=V_{B3right}=59.46$ V and $V_{B4left}=V_{B4right}=12.14$ V with respect to the second monochromator exit slit.

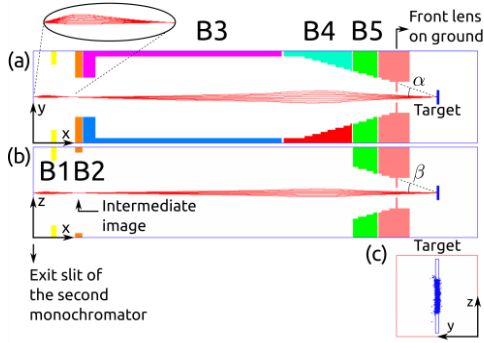

Fig. 5. Same as Fig. 3 for the case $\alpha_{mean} = 2°$. Electron trajectories are calculated for the following conditions: $V_{B1}$=-4.1 V, $V_{B2}$=12.14 V, $V_0$=59.46 V, $E_0$=60 eV, $V_{B5}$=59.46 V, $V_{B3left}$=59.28 V, $V_{B3right}$=59.64 V, $V_{B4left}$=12.21 V and $V_{B4right}$=11.99 V.

By applying deflection voltages of opposite sign to the split lenses B3 and B4 one can make the beam travelling along the optical axis, even when the mean exit angle of the monochromator forms angle of up to 3° with the optical axis. This is demonstrated with Fig. 4 and Fig. 5. Both figures display cross sections through the lens parallel to the dispersion plane of the monochromators ($\alpha$-plane / $xy$-plane) and perpendicular to it ($\beta$-plane / $xz$-plane) in panel (a) and (b), respectively. A bundle of electron trajectories leaving the center of the exit slit of the monochromator with Gaussian distributions of the angles $\alpha$ and $\beta$

$$P(\alpha) = \exp(-\alpha^2 / 2s_\alpha^2),$$
$$P(\beta) = \exp(-\beta^2 / 2s_\beta^2) \quad (8)$$

is shown as red lines. The variances are $s_\alpha$ = 0.94° and $s_\beta$ = 1.3°.

In Fig. 4, the distribution of electrons is centered around the optical axis of the transfer lens. As discussed in detail later, about 85% of all electrons leaving the monochromator fall into the field of view of the analyzer. Approximately the same holds for electrons leaving the monochromator at a mean angle of $\alpha_{mean} = 2°$ with respect to the optical axis (Fig. 5).

## IV. SIMULATIONS OF THE CATHODE EMISSION SYSTEM AND THE MONOCHROMATORS

As discussed above, simulations of the image formation at the target must involve the entire system from the cathode tip through the two sections of the monochromator and finally through the transfer lens. The distribution of angles at the target is determined by the distribution of angles in the beam produced by the cathode emission system (Eq. 3) and by the pass energies of the two monochromators. It is therefore indispensable to perform simulations including both the emission system and the two monochromators. Because of the high current load, the electron-electron repulsion must be included. The simulations are based on home-grown computer codes. The basic principles of these codes are described in Ref. [17]. Here we merely quote results specific to the system under consideration.

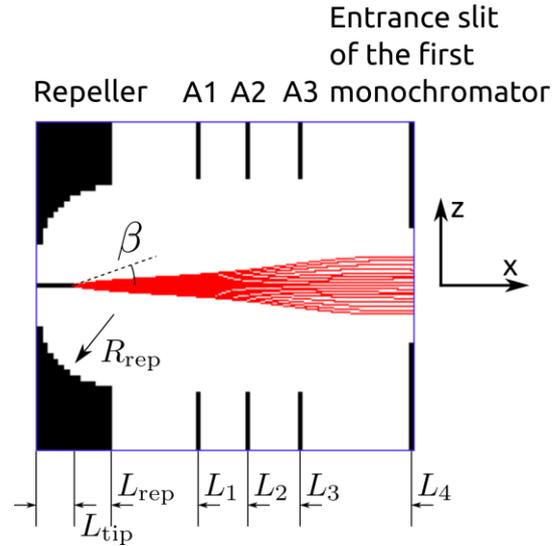

Fig. 6. $xz$-plane cross-section of the standard cathode emission system (system #24). The red lines show trajectories of electrons emerging from the cathode tip.

Because of the complexity of the simulations we need to focus on a particular setting of the monochromators. We choose deflection voltages of 8 V and 1 V for the first and second monochromator respectively, which yields an electron beam with a theoretical FWHM of about 2.3 meV.

Two versions of the emission system are available. One is used in most (including commercial) HREELS instruments (system #24

in Table 1). The other one has been developed recently in the context of high resolution spin wave spectroscopy (system #58 in Table 1) [7]. Due to its smaller dimensions the latter system yields higher currents, however, also a broader distribution of angles (Table 2). In the following, we discuss simulations of our electron source with both emission systems. Aiming at higher wave vector resolution (Eq. 3), we make use only of the standard emission system #24 for the performance test. The key properties of the electron beam leaving the two monochromators are shown in Fig. 7 as function of the input current. These are: The FWHM of the beam $\Delta E_\mathrm{M}^\mathrm{theo}$ (right side axis) and the "monochromatic" current (left axis). Also shown is the specific current defined as the current per energy interval at the peak of the monochromator transmission curve, which is at about 0.7 for each of the two monochromators. The transmission is less than unity because of the angular aberration of the deflector: Electrons having embarked on trajectories with larger off-axis angles appear at different, higher energies than electrons on the central trajectory. The specific current is therefore a useful quantity, serving to characterize the performance of the monochromator. With increasing input current more and more electrons are deflected from their path due to the increasing electron-electron repulsion.

| system # | $L_\mathrm{tip}$ | $L_\mathrm{rep}$ | $R_\mathrm{rep}$ | $L_1$ | $d_1$ | $L_2$ | $d_2$ | $L_3$ | $d_3$ | $L_4$ |
|---|---|---|---|---|---|---|---|---|---|---|
| 24 | 1.5 | 3.5 | 4.0 | 6.0 | 8.0 | 8.0 | 8.0 | 10 | 8.0 | 14.2 |
| 58 | 0.6 | 2.2 | 1.5 | 3.2 | 2 | 4.1 | 3.0 | 5.0 | 3.0 | 7.0 |

Table 1: Dimensions of the cathode emission systems shown in Fig. 6 (all in mm). *L*, $R_\mathrm{rep}$ and *d* stand for the distances with respect to the repeller onset, the radius of the repeller (rep), and the diameter of the circular openings of the aperture lenses A1, A2 and A3.

| system # | $V_\mathrm{rep}$/V | $V_\mathrm{A1}$/V | $V_\mathrm{A2}$/V | $V_\mathrm{A3}$/V | $I^\mathrm{emission}$(opt)/μA | $I^\mathrm{input}$/nA | $s_\alpha$/° | $s_\beta$/° |
|---|---|---|---|---|---|---|---|---|
| 24 | -4.8 | 58 | 9.8 | -4 | 0.9 | 50 | 0.6 | 0.6 |
| 58 | -4.8 | 58 | 10 | -0.5 | 2 | 500 | 1.3 | 1.4 |

Table. 2: Comparison of achievable feed currents into the entrance slit of the first monochromator with a deflection voltage of $\Delta V_\mathrm{mono1} = 8\,\mathrm{V}$ ($E_\mathrm{mono1}^\mathrm{pass} = 4.7\,\mathrm{eV}$) for the old and the new cathode emission systems. In order of the columns, the table contains the system number, the voltage of the repeller and of the lenses A1, A2, and A3, the optimum cathode emission current, the feed current into the (0.3×3) mm² entrance slit of the first monochromator and the Gaussian variances of the angle distribution of the trajectories.

The slope of the specific current vs. the input current decreases and eventually becomes negative (blue triangles in Fig. 7). Increasing the input current beyond the maximum of the specific current curve has initially little influence on $\Delta E_\mathrm{M}^\mathrm{theo}$. Instead, the additional electrons end up at higher energy, where the transmission vs. kinetic energy curve has a tail [13]. In order to avoid this tail, the monochromators should be fed with input currents at the maximum of the specific current or slightly below. Fig. 7 instructs us that these points of operation lie at 125 nA and 175 nA input current for the emission system #24 and #58, respectively. However, according to Table 2 the conventional emission system (#24) cannot provide 125 nA feed current suitable to run the monochromators at optimum performance. Because of the 50 nA limit (Table 2) the maximum monochromatic current for $\Delta E_\mathrm{M}^\mathrm{theo} = 2.3\,\mathrm{meV}$ is about 0.12 nA (magenta solid circles in Fig. 7).

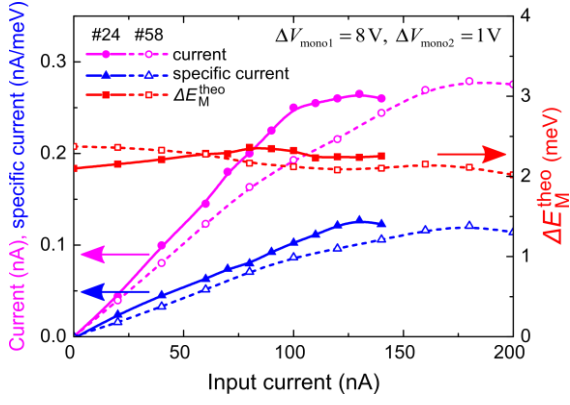

Fig. 7. Output current, specific current and FWHM of the electron source $\Delta E_\mathrm{M}^\mathrm{theo}$ as function of the input current for the systems #24 and #58 shown as solid and open symbols, respectively.

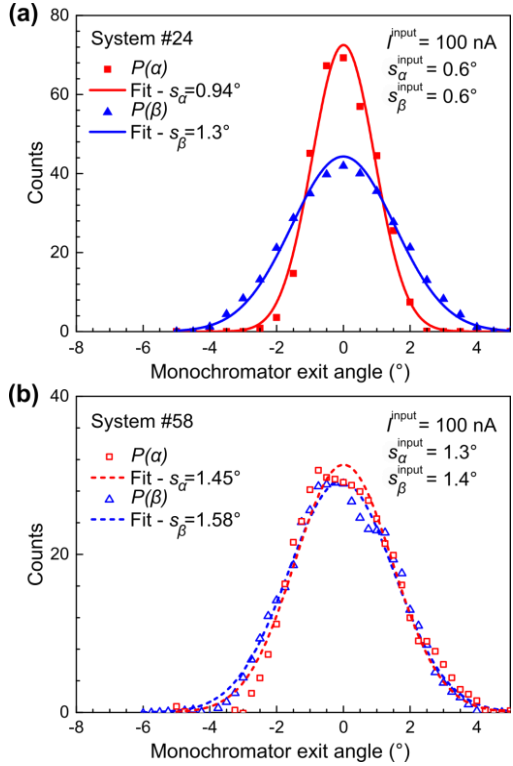

Fig. 8. Distributions of exit beam angles $P(\alpha)$ (squares) and $P(\beta)$ (triangles) for the double monochromator system when fed by cathode system #24 (a) and #58 (b). The input current is 100 nA. The variances of the input beam ($s_\alpha^\mathrm{input}$, $s_\beta^\mathrm{input}$) and the exit beam ($s_\alpha$ and $s_\beta$), the latter fitted by a Gaussian function (lines), are also indicated.

All data refer to (0.3×3) mm² entrance slits of the first and the second monochromators and to the (0.3×2) mm² exit slit of the second monochromator. The exit slit height is reduced to decrease the angular distribution in the $\beta$-plane, and thus to increase the momentum resolution.

The distributions of the electron exit angles $\alpha$ and $\beta$ out of the double monochromator system are shown in Fig. 8a and 8b for input beams provided by the emission system #24 and #58, respectively. The input current is 100 nA in both cases. Remarkably, the resulting width of the angle distribution is only marginally broader for system #58, despite the much larger angle width of the input beam. In particular the variance $s_\beta$, which eventually transforms into the wave vector resolution of the instrument, is nearly the same. The reason is that the first of the two monochromators not only acts as an energy filter, but also as an angular aperture: electrons having embarked on trajectories at larger angles appear in the high-energy tail of the transmission function of the first monochromator. The second monochromator filters out those electrons energy-wise and therefore also the large-angle trajectories.

## V. SIMULATIONS FOR THE TRANSFER LENS

The transfer lens systems was originally designed with the intention that B3 and B4 should be essentially at target potential with merely slight deflection voltages applied. The second focus at the target was to be achieved by applying a high potential to B5. While the lens system does operate under those conditions and produces a small spot on the target, it also causes a relatively broad distribution of angles $P(\alpha)$, $P(\beta)$ at the target. The resulting wave vector resolution is then too low to be useful for displaying dispersion curves at the output of the hemispherical analyzer. An alternative mode of operation, still involving an intermediate image, is therefore adopted, in which B5 is at target potential and B4 at lower potential. Then the cardinal plane shifts backwards from the center of B5 into the center of B4 (see Fig. 4). The image at the target becomes larger thereby and the angle distributions $P(\alpha)$, $P(\beta)$ become narrower. In the course of the simulation study it was found furthermore that there is a large redundancy in the operating potentials. We found that with no loss of performance we could couple B2 and B4 to the same potential and also B3 and B5 to the target potential (except for the deflection voltage between B3left and B3right and between B4left and B4right). Deflection voltages disregarded, one is then left with only two lens potentials to be optimized for each impact energy and monochromator energy.

In the simulations the transfer lens system is fed with the angle distribution of electrons $P(\alpha)$ and $P(\beta)$ with which they leave the monochromator exit slit (Fig. 8a). A random distribution over the starting positions inside the entrance slit is assumed, since also the cathode emission system provides a nearly even distribution of coordinates at the entrance slit of the monochromator. For the impact energy of $E_0$ = 60 eV the distribution of angles at the target is displayed in Fig. 9. The FWHM of $P(\beta)$ is 3.0° and 1.4° for $P(\alpha)$. This is smaller than 4.05° and 2.7 for $\beta$ and $\alpha$-angles, respectively, deduced from phase-space conservation considerations (Sec. II). The reason is that the transmission of all optical elements is reduced for larger angles.

The distribution in $\beta$ transforms into the wave vector resolution of the instrument, assuming $\theta_i = \theta_f = 45°$, according to

$$\Delta q_{\parallel} = k_i \sin(45° + \Delta\beta/2) - k_f \sin(45° - \Delta\beta/2). \quad (9)$$

For small losses and $E_0$ = 60 eV one obtains a wave vector resolution (FWHM) of $\Delta q_{\parallel}$ = 1.47 nm$^{-1}$. This is about 8% of the total wave vector range probed by the analyzer at $E_0$ = 60 eV (17.7 nm$^{-1}$ using Eq. 6 and 7).

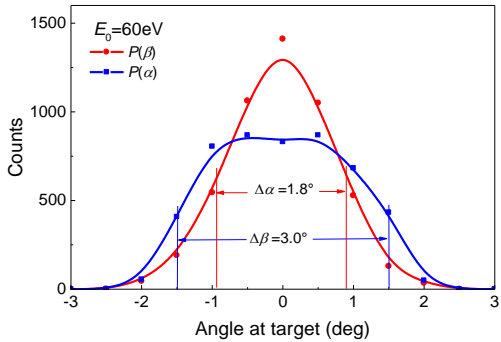

Fig. 9. Distribution of beam angles at the target $P(\alpha)$, $P(\beta)$. The wave vector resolution is determined by the FWHM of $P(\beta)$. The lines are spline-fits as guides to the eye.

Fig. 10 shows the key properties of the lens as a function of the electron energy at the target. Note that the electron energy at the target $E_0$ and the target voltage $V_0$ are related by $E_0 = eV_0 + E_{mono2}^{pass}$. The fraction of electrons ending in the intended (0.04×1) mm$^2$ target area, denoted as *transmission*, is shown in Fig. 10a; the FWHM of $P(\beta)$, denoted as $\Delta\beta$, in Fig. 10b and $\Delta q_{\parallel}$, in Fig. 10c. The solid lines in Fig. 10b and c are fits by a simple power law, which for $\Delta q_{\parallel}$ is

$$\Delta q_{\parallel} = 0.1987 \times (V_0/V)^{0.4086} \text{ nm}^{-1}. \quad (10)$$

The potentials on lens B1/B2 and B4 can be parameterized as function of the voltage at the target $V_0$. The data in Fig. 10 are obtained using the lens table

$$V_{B1} = -0.93V - 0.051 \times V_0 \text{ and} \quad (11)$$

$$V_{B2} = V_{B4} = -0.50V + 0.21 \times V_0. \quad (12)$$

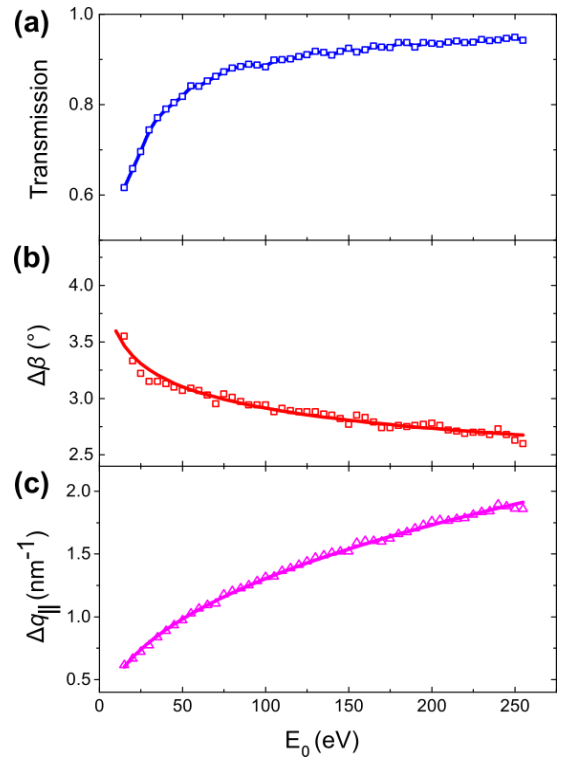

Fig. 10. (a) Fraction of electrons reaching the intended (0.04×1) mm$^2$ target area (*transmission*). (b) FWHM of $P(\beta)$, denoted as $\Delta\beta$. (c) FWHM of the momentum resolution, denoted as $\Delta q_{\parallel}$ (see Eq. 9).

Fig. 11 shows that the simulated optimum lens table agrees with the optimum found experimentally. These data refer to $E_{mono2}^{pass} = 0.54\text{eV}$ (i.e. deflection voltage $\Delta V_{mono2} = 1.0\text{V}$). For small variations in the deflection voltage only the voltage $V_{B1}$ needs to be adjusted. A generalization of Eq. 11 that works well for monochromator deflection voltages between $\Delta V_{mono2} = 0.2\,\text{V}$ and $\Delta V_{mono2} = 1.2\,\text{V}$ is

$$V_{B1} = 1.57\,\text{V} - 2.5 \times \Delta V_{mono2} - 0.051 \times V_0. \quad (13)$$

## VI. OPTIMIZATION ROUTINES

The practical use of the instrument as described above rests with the efficiency by which one is able to find a set of optimum voltages for all optical elements. This is a nontrivial task, since the performance of the instrument depends literally on all voltages. For example, we have seen in the simulations that the angular distribution of the beam delivered by the cathode emission system transfers into the angle distribution of the beam at the target, and from there into the intensity and energy resolution of the entire system. Furthermore, all voltages depend on the required energy resolution and the desired impact energy at the target. The optimum voltages are even sensitive to the heating current of the cathode. A variation of that current by merely 0.01 A already requires an adjustment of most other voltages.

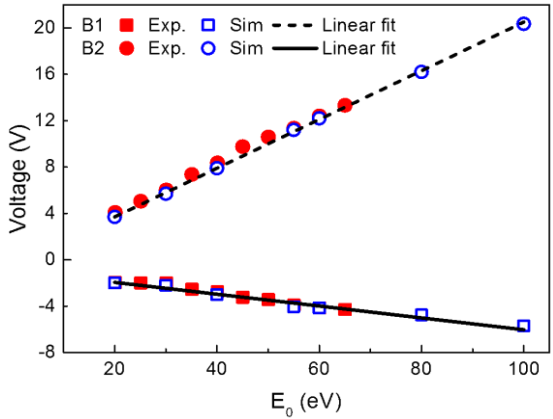

Fig. 11. Lens table for the transfer lens. Experimentally found optimum voltages for $V_{B1}$ and $V_{B2} = V_{B4}$ are shown as solid red squares and circles, respectively, with $V_{B3} = V_{B5} = V_0$. Optimum voltages obtained with the electron optical calculations are shown as open blue squares and circles, respectively. The dashed and solid magenta lines are linear fits according to Eq. 11 and 12. All data refer to $E_{mono2}^{pass} = 0.54\,\text{eV}$.

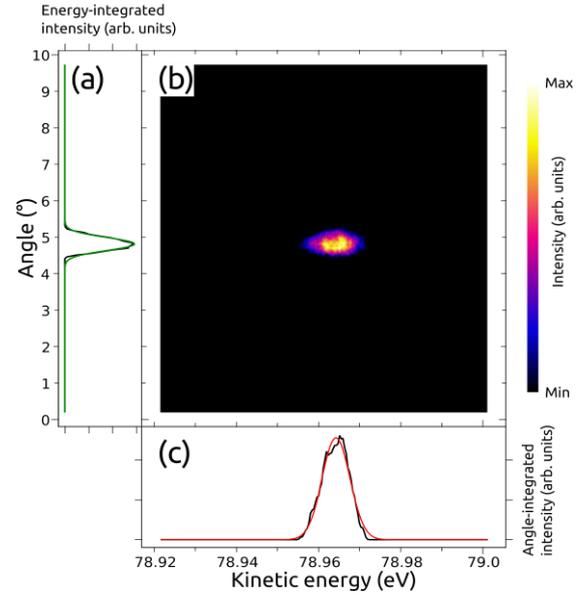

Fig. 12. Example of a live detector image showing the electron beam in the specular direction measured with the MCP voltage of 1120 V. (a) Intensity integrated over kinetic energies. (b) Snapshot of a detector image. (c) Intensity integrated over angles. Black lines in (a) and (c) correspond to integrated intensity profiles and green and red to corresponding fits by Gaussians.

It is therefore essential to have automated routines for finding the optimum voltages. For conventional HREELS, only the integrated intensity of the specular beam in a narrow range of angles and energies can be used for optimization. In the present case, energy and angle resolution are available simultaneously. Therefore, a live analysis of both resolutions and the total intensity is possible, and the quantity to be optimized can be any arbitrary combination of these three quantities. The monochromatic current at the sample scales approximately with the second power of $\Delta E_{total}^{exp}$. It is therefore meaningful to optimize the ratio of the intensity at the channel plate over the square of $\Delta E_{total}^{exp}$. A characteristic image of the elastic beam during optimization is shown in Fig. 12b. The figure shows the intensity spot of the optical readout of the hemispherical analyzer MCP after specular reflection of the primary electron beam from a clean Cu(111) sample. During the optimization, the data are taken with a low voltage on the MCP to prevent possible damage due to the high intensity of the specular beam. It is in fact important to implement a "watchdog" system to automatically reduce the voltage on the MCP within microseconds in case of the detected beam intensity exceeding the preset threshold limit. It is worth noting that by reducing the voltage on the MCP, the observed intensity is not linear with the observed electron beam intensity. In this case, the MCP counts events only when more than a single electron arrives within the time span required to form the electron cloud by multiplication, thus producing a measured specular beam narrower than the actual one.

Fig. 12a and c display the energy-integrated intensity as a function of angle and the angle-integrated intensity as a function of energy, respectively. The green and the red lines are Gaussians fitted to the integrated data from which the energy resolution $\Delta E_{total}^{exp}$ or the angular resolution can be determined. The optimization routine calculates the performance measure $O = I_{total} / (\Delta E_{total}^{exp})^2$ in which $I_{total}$ is the integrated intensity of the spot. $O$ is then maximized by sequential variation of all voltages in the cathode emission system, the monochromator section and the transfer lens. Only the deflection voltages of the monochromator, the analyzer settings and the cathode heating current are kept constant. Note that the definition of $O$ is flexible and can be adjusted to match any specific need (for instance, higher intensity with poorer energy resolution, or momentum resolution). In addition to the single voltage optimization one may also make use of a pairwise optimization. This is particularly useful for pairs like the mean voltage of the inner and outer monochromator deflection plates versus the mean voltage on the upper and lower monochromator deflection plates. Once a complete set of optimum voltages has been obtained, one may also vary the cathode heating current and repeat the optimization procedure to find the absolute maximum.

Once a set of optimum voltages is determined, it is stored and can be recovered when needed. The reproducibility of an optimized setting is such that only a brief optimization is needed for a new experiment. In most cases it is then enough to run shortened routines, such as cathode mean potential versus first monochromator mean potential, and a second routine involving the voltages of the transfer lens.

**VII. PERFORMANCE TEST**

To illustrate the performance of the complete instrument, we measured the well-known dispersion of the surface phonons on the Cu(111) surface in the $\overline{\Gamma M}([1\overline{1}0])$-direction. The surface is cleaned by cycles of Ar ion sputter (800 V) and annealing (500°C). In Fig. 13, we present the data obtained in 7 minutes acquisition time. The left panel shows the intensity map displayed in a two-dimensional energy vs. wave vector plot. The impact energy $E_0$ is 112 eV. Wave vectors range from 0.4 to 21.1 nm$^{-1}$, well beyond the $\overline{M}$ point of the surface Brillouin zone at 14.2 nm$^{-1}$. The red lines are the dispersion curves as calculated by density functional perturbation theory [18].

The panel on the right displays the intensity integrated between 8 nm$^{-1}$ and 9 nm$^{-1}$, indicated by dotted lines in the left panel. The peaks can be assigned as follows: at ±11.66 meV the energy gain and loss from the Rayleigh phonon is observed. The elastic diffuse scattering (caused by disorder of the surface lattice) is detected at 0 meV. Finally, a second energy loss due to a surface resonance ($S_2$ and $S_2$')[18, 19] is registered at ±26.8 meV. All peaks sit on a multi-phonon background. This background increases with the electron impact energy, as reported, e.g., for Cu(111)[20] and for Ni(100)[21]. Shape and magnitude of the multi-phonon background can be estimated by making use of the fact that the multi-phonon background is a smooth function, that there are no phonons below the lowest surface mode, i.e. the Rayleigh mode, and finally that there are no single phonon losses beyond the surface resonance $S_2$ and $S_2$' at ±26.8 meV [18]. With the multi-phonon background subtracted, the elastic diffuse peak has a FWHM of 3.9 meV, which is the resolution of the complete set-up $\Delta E_{total}^{exp}$, including the sample.

The experimental resolution in Fig. 13 compares well with the predicted resolution of the instrument. The total resolution of the electron source $\Delta E_M^{theo}$ with two monochromators in sequences is [13]

$$\Delta E_M^{theo} = \frac{\Delta E_{mono1}^{theo} \Delta E_{mono2}^{theo}}{\sqrt{(\Delta E_{mono1}^{theo})^2 + (\Delta E_{mono2}^{theo})^2}} \quad (14)$$

where $\Delta E_{mono1}^{theo}$ and $\Delta E_{mono2}^{theo}$ are the resolution of the first and second monochromators, respectively.

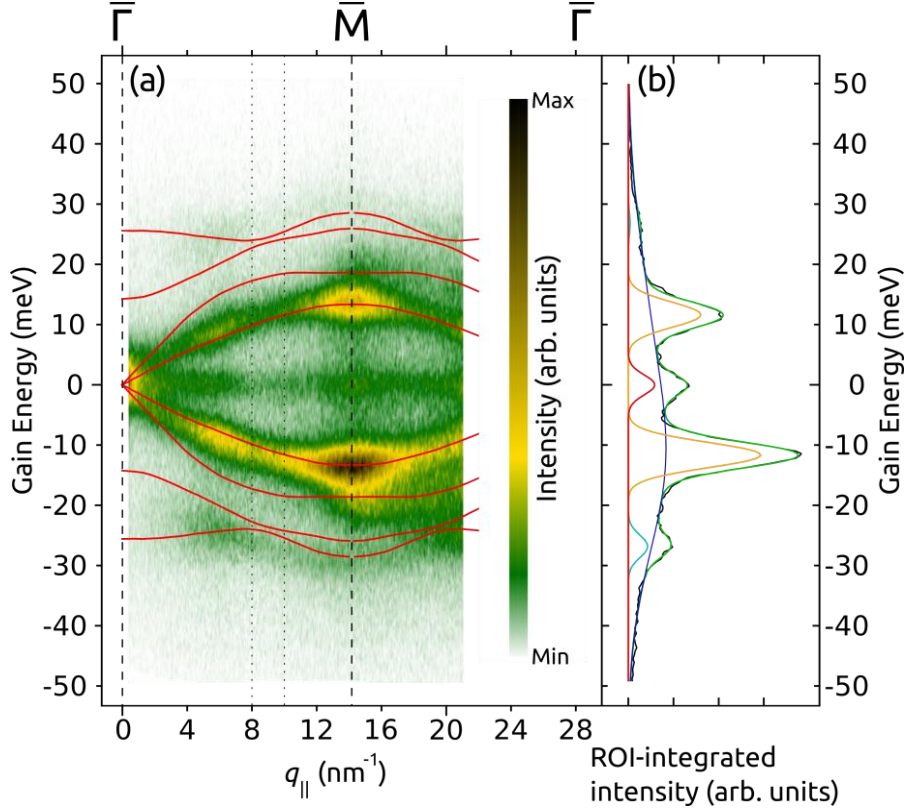

Fig. 13: (a) Intensity map of the inelastic electron scattering from the Cu(111) surface in the $\overline{\Gamma}\,\overline{\mathrm{M}}$ direction. The red lines correspond to the surface phonon dispersion bands calculated by density functional perturbation theory [18]. (b) Intensity of the inelastic electron scattering integrated in the range of (8-9) nm$^{-1}$ (see text for details).

The overall resolution of the instrument is then calculated as the width of the convolution of two Gaussian transfer functions, one for the electron source, the other for the analyzer

$$\Delta E_{\mathrm{total}}^{\mathrm{theo}} = \sqrt{(\Delta E_{\mathrm{M}}^{\mathrm{theo}})^2 + (\Delta E_{\mathrm{A}}^{\mathrm{theo}})^2} \qquad (15)$$

Using Eq. 1 and 14 and the deflection voltages $\Delta V_{\mathrm{mono1}} = 8.1\,\mathrm{V}$ and $\Delta V_{\mathrm{mono2}} = 0.83\,\mathrm{V}$, one calculates the total resolution of the electron source to be $\Delta E_{\mathrm{M}}^{\mathrm{theo}} = 2.06\,\mathrm{meV}$. According to the factory tests of the used Scienta R4000 analyzer for the settings used here $\Delta E_{\mathrm{A}}^{\mathrm{theo}}$ equals 3.3 meV, yielding a calculated total energy resolution $\Delta E_{\mathrm{total}}^{\mathrm{theo}}$ of 3.89 meV (Eq. 15). This is in excellent agreement with the measured $\Delta E_{\mathrm{total}}^{\mathrm{exp}}$ value of 3.9 meV.

## VIII. CONCLUSION

We have designed a monochromatic electron source producing an electron beam with tunable electron energy and an energy resolution of a few meV. The electron source equipped with the transfer lens is designed such that it can be readily bolted onto existing UHV systems for photoemission spectroscopy in a similar way as ultraviolet and x-ray radiation sources. With the two-dimensional readout of modern hemispherical analyzers a large multiplex gain is achieved compared to conventional spectrometers used for electron energy loss spectroscopy.

## ACKNOWLEDGMENTS
F.C.B. acknowledges financial support from the Initiative and Networking Fund of the Helmholtz Association, Postdoc Programme VH-PD-025. The authors thank J. Åhlund, M. Lundqvist and R. Moberg for fruitful discussion as well as A. Franken, C. Elsässer, H. Stollwerk, U. Viehhöver W. Hürttlen and M. Wirde for their excellent technical support.